\documentclass[pra,aps,superscriptaddress,amsmath,amssymb, nobibnotes,nofootinbib,twocolumn,10pt]{revtex4-2}
\usepackage{graphicx} 
\usepackage{amsthm}

\usepackage[citecolor=blue,pdfa=true,linktocpage=true,urlcolor=blue,colorlinks=true]{hyperref}
\usepackage{tikz}

\usepackage{soul}
\usepackage{multirow}

\DeclareMathOperator{\Tr}{Tr}
\newcommand{\nc}{\newcommand}
\nc{\n}{\textrm{n}}
\usepackage{xcolor}

\usepackage{braket}

\newcommand{\ot}{\otimes}

\def\pra{Phys. Rev. A}
\def\prl{Phys. Rev. Lett.}
\def\physrep{Phys. Rep.}

\begin{document}

\title{Separable ellipsoids around  multipartite states}

\author{Robin Y. Wen}
\affiliation{\it California Institute of Technology, 1200 E. California Boulevard, Pasadena, California
91125, USA}

\author{Gilles Parez}
\affiliation{\it Laboratoire d’Annecy-le-Vieux de Physique Théorique (LAPTh), CNRS, Université Savoie Mont Blanc, 74941 Annecy, France}

\author{Liuke Lyu}
\author{William Witczak-Krempa}
\affiliation{\it D\'epartement de Physique, Universit\'e de Montr\'eal, Montr\'eal, QC H3C 3J7, Canada}
\affiliation{\it Centre de Recherches Math\'ematiques, Universit\'e de Montr\'eal, Montr\'eal, QC H3C 3J7, Canada}

\affiliation{\it Institut Courtois, Universit\'e de Montr\'eal, Montr\'eal, QC H2V 0B3, Canada}

\author{Achim Kempf}
\affiliation{\it Department of Applied Mathematics, University of Waterloo, Waterloo, ON N2L 3G1, Canada}
\affiliation{\it Department of Physics, University of Waterloo, Waterloo, ON N2L 3G1, Canada}
\affiliation{\it Institute for Quantum Computing, University of Waterloo, Waterloo, ON N2L 3G1, Canada}

\begin{abstract}
We show that, in finite dimensions, around any $m$-partite product state $\rho_{\rm prod}=\rho_1\otimes...\otimes\rho_m$, there exists an ellipsoid of separable states centered around $\rho_{\rm prod}$. This separable ellipsoid contains the separable ball proposed in previous works, and the volume of the ellipsoid is typically exponentially larger than that of the ball, due to the hierarchy of eigenvalues in typical states. We further generalize this ellipsoidal criterion to a trace formula that yields separable region around all separable states, and further study biseparability. Our criteria not only help numerical procedures to rigorously detect separability, but they also lead to a nested hierarchy of SLOCC-stable subsets that cover the separable set. We apply the procedure for separability detection to 3-qubit X states, genuinely entangled 4-qubit states mixed with noise, and the 1d transverse field Ising model at finite temperature to illustrate the power of our procedure for understanding entanglement in physical systems.
\end{abstract}

\maketitle

\section{Introduction}

Entanglement plays a key role in quantum computing \cite{10NielsenChuangText,2010LaddQCReview,17BoyerDQC1,18KlcoDigitizedSFQC}, communication \cite{10YuanPhotonQCCommR,18StephanieQI} and  sensing \cite{17DegenQS,17RosskpfQS,16HuntemannIonClock}. Moreover, is it a prominent tool in condensed matter systems \cite{16Laflorencie} for probing complex many-body phenomena, such as quantum phase transitions \cite{OAFF02, ON02, VLRK03, CC04} and topological phases~\cite{kitaev2006topological, levin2006detecting}. It is therefore important to distinguish the occurrence or absence of entanglement among quantum states. We here investigate this problem for systems composed of $m$ subsystems with finite-dimensional Hilbert spaces $\mathcal{H}_{i}$, with $i=1,..., m$. We recall that a state $\rho$ acting on $\mathcal{H}_{1}\otimes...\otimes\mathcal{H}_{m}$ is called unentangled, i.e., separable\footnote{One can similarly define separability for all Hermitian matrices (un-normalized states). See \cite{02GurvitsLSBaMMBS} for example.}, if and only if there exist density matrices $\rho_{i,k}$  
acting on $\mathcal{H}_{i}$ such that $\rho=\sum_{k}{p_k}\rho_{1,k}\ot...\ot\rho_{m,k}$, where $\sum_{k}{p_k}=1$ and $p_k~\geqslant~0$ for all $k$. Despite the apparent simplicity of the condition of separability, determining whether a given state is entangled or separable is in general NP-hard~\cite{03Gurvitis_Complexity,08GharibianNP}. The challenge becomes even more pronounced when dealing with multipartite cases \cite{19CunhaTripartiteEntanglement,16BengtssonMulti-partite}.

A particularly interesting aspect of the separability problem is the characterization of separable balls (hypersphere) in the space of quantum states. Indeed, the set of all separable states is convex, and a key aspect of the geometry of a convex set is the size of the largest ball that fits inside. Refs.~\cite{98Zyczkowski_Separable_Ball,99Vidal,99Braunstein_Separable_Mutipartite,01Rungta_Qudit,03Gurvits_Multipartite_Mixed} showed the existence of a separable ball around the maximally mixed states, $\frac{1}{D} \mathbb{I}_1\otimes...\otimes \mathbb{I}_m$, where $D$ is the total dimension of the system, while providing successively better lower bounds for the radius. Ref.~\cite{02GurvitsLSBaMMBS} found the exact size of the separable ball in the Frobenius norm for the bipartite case, but the exact size of the ball has not yet been established for the generic multipartite case.

Based on these results, recent works have established the existence and possible sizes of separable balls around other bipartite or multipartite states of interest, such as product states of the form $\rho_{\rm prod}=\rho_1\otimes...\otimes\rho_m$. Ref.~\cite{16LamiEntanglementSaving} first showed the existence of a separable ball around any full-rank bipartite product state, and Refs.~\cite{20Hanson_Markovian,23Wen_Ball} found a lower bound on the radius of the separable ball around the multipartite product states to be $2^{1-m/2}\lambda_{\rm min}(\rho_{\rm prod})$, which is proportional to the smallest eigenvalue of $\rho_{\rm prod}$. 

The existence of separable balls has important implications for the structure of entanglement in quantum many-body systems \cite{23Wen_Ball,24Parez}. In particular, any quantum system starting with a full-rank product state will remain unentangled for a finite amount of time regardless of the dynamics~\cite{23Wen_Ball}. 
Ref.~\cite{24Parez} used separable balls to argue that multipartite entanglement typically dies during the generalized evolution of a quantum state, including in finite time, distance or temperature. However, the examples of Ref.~\cite{24Parez} demonstrated that the ball criterion of Ref.~\cite{23Wen_Ball} is far from optimal: states are found to be separable well before entering the separable ball. This motivates further work to improve the ball criterion. 

In this work, we show that the separable region around any full-rank multipartite product state contains an ellipsoid. This separable ellipsoid uses all the eigenvalues of the product states, instead of just the minimal one for the separable ball. We find that 
the volume of the ellipsoids is exponentially larger than that of the balls considered in Refs.~\cite{20Hanson_Markovian,23Wen_Ball} owing to the typically large hierarchy of eigenvalues density matrices. For instance, it was shown~\cite{Majumdar} that random density matrices have an ensemble average $\langle\lambda_{\rm min}\rangle\propto 1/D^3$. Furthermore, for product states that are not full-rank, lower-dimensional separable ellipsoids naturally emerge from the full-rank subspaces of the states. Using a scaling relation for the separable ellipsoid, we give a new sufficient criterion for multipartite separability based on trace (Eq.~\eqref{eq:trace-condition}). We first generalize this criterion to describe separable regions around any multipartite states, the \emph{General Trace Criterion (GTC)}, which serves as a powerful and rigorous criterion for detecting separability when combined with simple numerical procedures. We next generalize to biseparability and characterize a biseparable region around any biseparable state. On one hand the GTC yields a hierarchy of non-convex subsets that are stable under stochastic local operations and classical communication (SLOCC)~\cite{Dur2000}, and cover the interior of the separable space. On the other hand, our benchmarks on 3 and 4 qubit states show that the GTC can produce cutting-edge outcomes for separability detection, and we employ it to obtain new results regarding the 1d Ising model at finite temperature, showing the convergence of separability thresholds with those from the positive partial trace (PPT) criterion~\cite{Peres1996,Horodecki1996}.

\section{Separable Ellipsoid}
For an $m$-partite quantum system, we denote the radius of the separable ball around the maximally mixed state $\frac{1}{D}\mathbb{I}$ by $\frac{1}{D}c_m$. Here $c_m$ is the dimensionless factor that controls the size of the ball, and the distance is measured with the Frobenius norm, $\lVert X\rVert_{\rm F} = \sqrt{\Tr(X^\dagger X)}$. The baseline result for $c_m$ is established in Ref.~\cite{03Gurvits_Multipartite_Mixed} to be at least $2^{1-m/2}$, which is optimal in the bipartite case~\cite{02GurvitsLSBaMMBS}. For $m$-partite quantum systems ($m\geqslant 3$) with each subsystem having the same dimension $d$, to our knowledge the best lower bound for $c_m$ is
\begin{equation}
c_m\geqslant\left\{
\begin{array}{ll}
      \sqrt{\frac{54}{17}}\times (\frac{2}{3})^{\frac{m}{2}}, & d=2, \\
      \sqrt{\frac{d^m}{(2d-1)^{m-2}(d^2-1)+1}}, & d\geqslant 3 ,\\
\end{array} 
\right.
\label{eq:cm}
\end{equation}
which was shown  for the $m$ qubits \cite{07Hildebrand_multiqubit} and qudits \cite{05Gurvits_better_bound}. 

{\it Theorem 1}: Consider arbitrary positive definite Hermitian operators $\rho_i$, $1\leqslant i\leqslant m$, acting on finite-dimensional Hilbert spaces $\mathcal{H}_{i}$. Define $\rho_{\rm prod}:=\rho_1\ot...\ot \rho_m$. If a Hermitian operator $\rho$ acting on $\mathcal{H}_{1}\otimes...\otimes \mathcal{H}_{m}$ satisfies $\lVert \rho_{\rm prod}^{-1/2}\rho\rho_{\rm prod}^{-1/2}-\mathbb{I}\lVert_{\rm F}\leqslant c_m$, then $\rho$ is separable. 

{\it Proof}:  Let $C:=\rho-\rho_{\rm prod}$ and 
\begin{equation}
    \Delta:= \rho_{\rm prod}^{-\frac{1}{2}}C \rho_{\rm prod}^{-\frac{1}{2}} = \rho_{\rm prod}^{-\frac{1}{2}}\rho\rho_{\rm prod}^{-\frac{1}{2}}-\mathbb{I}\nonumber\label{eq:Delta}.
\end{equation}
With the theorem assumption, we have $\lVert\Delta\lVert_{\rm F}\leqslant c_m$, so we know that $\mathbb{I}+\Delta$ is separable \cite{02GurvitsLSBaMMBS,05Gurvits_better_bound,07Hildebrand_multiqubit}. We then have that
\begin{align}
    \rho_{\rm prod}^{\frac{1}{2}}(\mathbb{I}+\Delta)\rho_{\rm prod}^{\frac{1}{2}}=\rho_{\rm prod}+C=\rho\nonumber
\end{align}
is also separable, since $\rho_{\rm prod}^{\frac 12}$ is SLOCC, being an invertible, local operator (thus preserving separability) \cite{21Qian}. \hfill\qedsymbol{}

\begin{figure}
    \centering
    \includegraphics[width=0.21\textwidth]{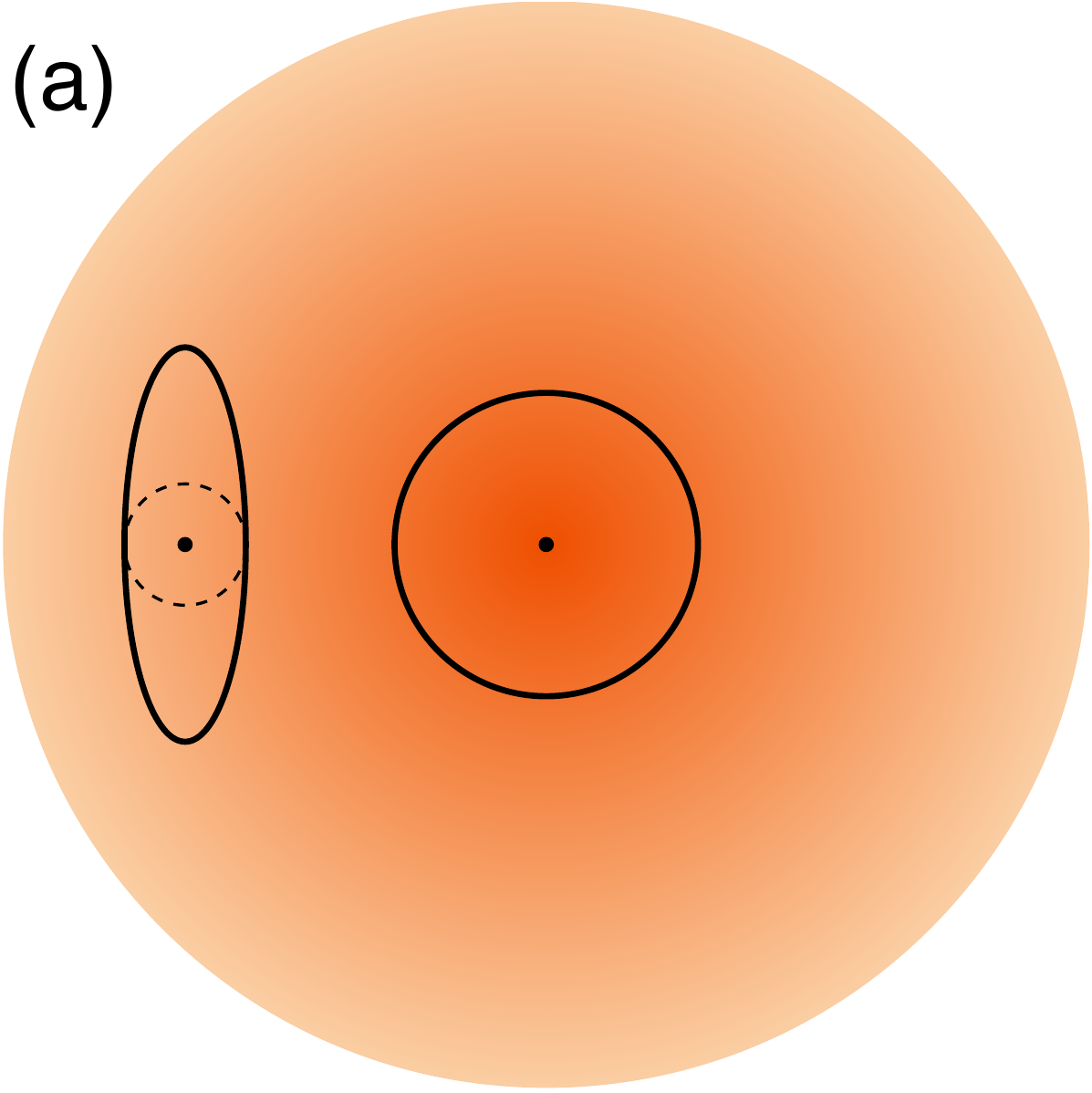}
    \includegraphics[width=0.243\textwidth]{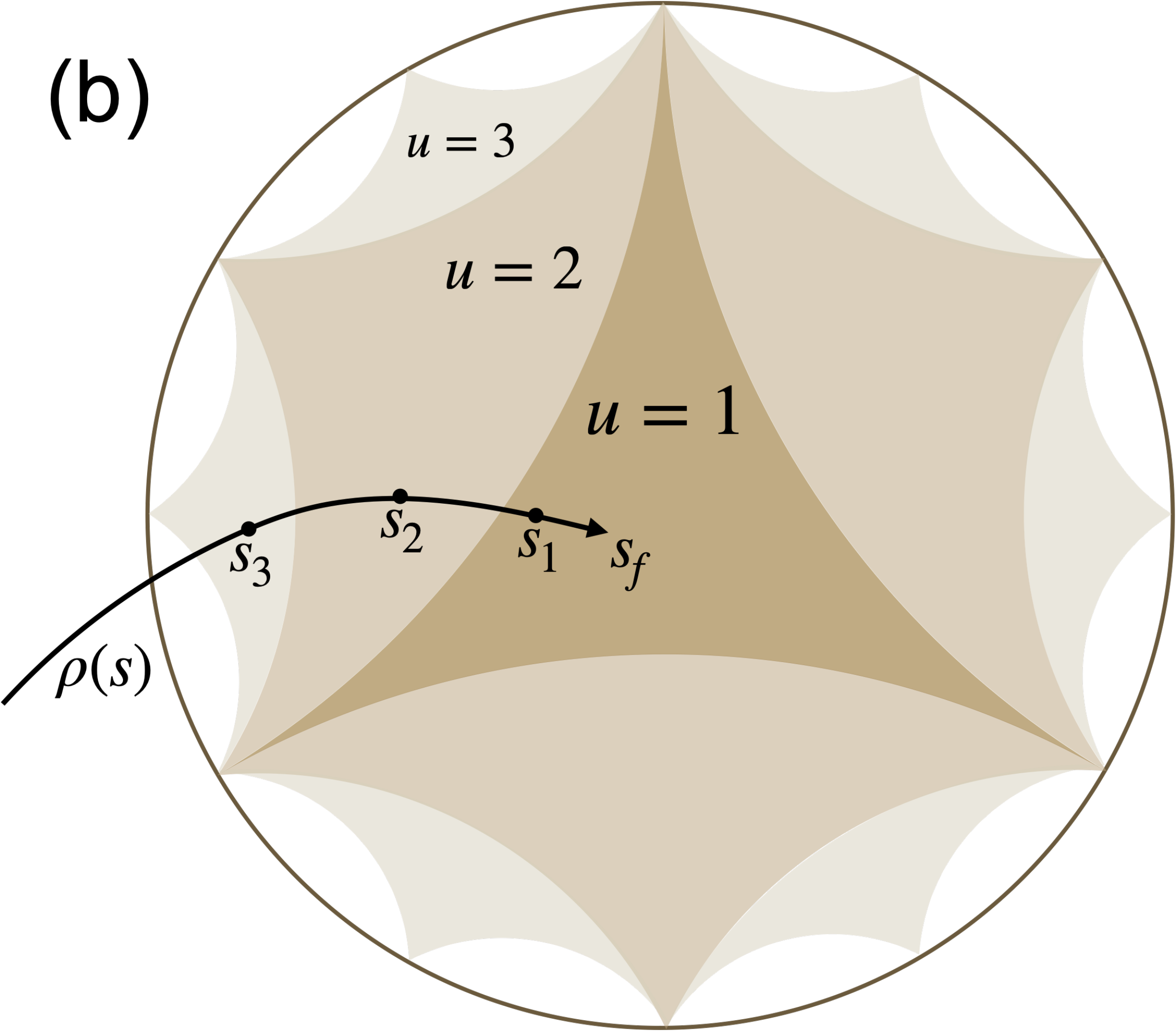}
    \caption{(a) Illustration of the separable balls and ellipsoids around separable states. The separable ellipsoid around the fully-mixed state in the center of the separable region coincides with the largest separable ball. For states away from the center, the separable ball (dotted lines) is the largest one in the ellipsoid (solid line). At the boundary, the ellipsoid has a lower dimension. (b) Hierarchy given by the generalized trace criterion (GTC). The circle represents the separable set, approximated by subsets \(\mathcal{S}_u\) corresponding to the \(u\)-component GTC. A curved line represents the evolution of a state $\rho(s)$ parametrized by $s$. Intermediate markers \(s_i\) denote states certified by the $i$-component GTC.}
    \label{fig:ellipsoid}
\end{figure}

The separable region characterized by the above theorem is an ellipsoid. To find the center and the lengths of the semi-axes of the ellipsoid, we use the diagonal basis of $\rho_{\rm prod}$ such that $\rho_{\rm prod}=\sum_{i,j}\lambda_{i}\delta_{ij}\ket{\lambda_i}\bra{\lambda_j}$ and $\rho=\sum_{i,j}\rho_{ij}\ket{\lambda_i}\bra{\lambda_j}$. Then the inequality $\lVert \Delta\lVert_{\rm F}\leqslant c_m$ can be equivalently expressed as
\begin{align}
    \lVert \Delta\rVert_{\rm F}^2&=\sum_{i,j}\left|\frac{\rho_{ij}-\delta_{ij}\lambda_{i}}{\sqrt{\lambda_{i}\lambda_{j}}}\right|^2\leqslant c_m^2.
    \label{eq:ellipsoid-condition}
\end{align}
Therefore, in the diagonal basis, the separable ellipsoid is centered at $\delta_{ij}\lambda_{i}$ with $c_m\sqrt{\lambda_{i}\lambda_{j}}$ as the length of the semi-axis for each $i,j$. The shape of the ellipsoid depends on the eigenvalues of $\rho_{\rm prod}$ and the direction of the ellipsoid depends on the corresponding eigenvectors. The separability criterion~\cite{23Wen_Ball} $\lVert \rho-\rho_{\rm prod}\lVert_{\rm F}\leqslant c_m\lambda_{\rm min}(\rho_{\rm prod})$ corresponds to the biggest ball inscribed in this ellipsoid, which we give a schematic illustration in (a) of Fig.~\ref{fig:ellipsoid}.

\subsection{Volume improvement}
We now discuss the improvement obtained by considering the ellipsoid around the full-rank product states instead of the ball. A large hierarchy of eigenvalues naturally occurs in physical states, such as reduced density matrices (RDMs) coming from local Hamiltonians, leading to a very small $\lambda_{\rm min}(\rho)$, which makes the previous separable ball criterion less effective. We can use Eq.~\eqref{eq:ellipsoid-condition} to quantify the volume of the separable ellipsoid. Let us consider a product state with eigenvalues $\lambda_1\geqslant\lambda_2\geqslant\cdots\geqslant\lambda_D=\lambda_{\rm min}$. The ratio of volumes is then~\cite{98Zyczkowski_Separable_Ball}
\begin{align}
\label{eq:R}
   \mathcal R\equiv \frac{\mbox{vol(elli.)}}{\mbox{vol(ball)}} = \left(\frac{\lambda_1}{\lambda_{\rm min}}\right)^{\! D}\left(\frac{\lambda_2}{\lambda_{\rm min}}\right)^{\! D}\cdots \left(\frac{\lambda_{D-1}}{\lambda_{\rm min}}\right)^{\! D},
\end{align}
where we have not imposed the normalization constraint for states belonging to the respective volumes, but this makes little difference as $D\geqslant 4$.

To get a sense of the scales involved in physical systems, let us consider 3 adjacent spins in the ground state of the 1d quantum Ising model in a transverse field~\cite{1970Pfeuty_TIM,15Dutta_QPT_TIM}:
\begin{align}
    H=-\sum_{i=-\infty}^\infty(X_i X_{i+1} -h Z_i)\,
    \label{eq:1D-Ising}
\end{align}
The 3-spin state becomes separable at modest separations~\cite{Hofman2014}, and a simple target state to apply the ellipsoid criterion is $\rho_{\rm prod}=\rho_1^{\otimes 3}$, where $\rho_1$ is the RDM of a single site. At the quantum critical phase transition point, the transverse field is $h=1$, and the eigenvalues of $\rho_1$ read $\tfrac12\pm \tfrac1\pi$. The eigenvalues of $\rho_{\rm prod}$ are thus $(0.548,0.122,0.027,0.006)$ with the middle two being triply degenerate, which leads to $\lambda_1/\lambda_{\rm min}\approx 91$, and an ellipsoid to ball ratio of $\mathcal R^{1-1/D^2}\approx 10^{62}$. When the transverse field takes the value $h=3$ instead, we get a volume ratio of $10^{174}$, and the ratio further diverges as $h\to\infty$. It shows how naturally occurring eigenvalue hierarchies lead to exponential volume improvements with the ellipsoid. Further, the hierarchy will be even more pronounced for bigger subregions.

\subsection{Trace criterion}
Expressing the Frobenius norm condition in {\it Theorem 1} in terms of trace and using the cyclic property thereof,
we find
\begin{align}
\Tr[(\rho\rho_{\rm prod}^{-1})^2]-2\Tr[\rho\rho_{\rm prod}^{-1}]\leqslant c_m^2-D.
\label{eq:BATr}
\end{align} 
To improve the separability condition, we multiply $\rho$ by a coefficient $\alpha$, and find the optimal value such that $\alpha \rho$ satisfies Eq.~\eqref{eq:BATr}. Scaling with $\alpha$, the above inequality becomes
$\alpha^2\Tr[(\rho\rho_{\rm prod}^{-1})^2]-2\alpha\Tr[\rho\rho_{\rm prod}^{-1}]\leqslant c_m^2-D$. Minimization yields $\alpha=\Tr[\rho\rho_{\rm prod}^{-1}]/\Tr[(\rho\rho_{\rm prod}^{-1})^2]$, and the separability condition becomes 
\begin{equation}
\frac{\Tr[(\rho\rho_{\rm prod}^{-1})^2]}{\Tr[\rho\rho_{\rm prod}^{-1}]^2} \leqslant \frac{1}{D-c_m^2}.
\label{eq:trace-condition} 
\end{equation}
This is a sufficient condition for the separability of $\alpha \rho$, and hence of $\rho$, and the scaling relation improves {\it Theorem 1} by deforming the ellipsoid. In the case where $\rho_{\rm prod}$ is the maximally mixed state $\frac{1}{D}\mathbb{I}$, Eq.~\eqref{eq:trace-condition} reduces to $\Tr[\rho^2]\leqslant 1/(D-c_m^2)$, which was found in Ref.~\cite{03Gurvits_Multipartite_Mixed}. This criterion significantly improves over the separability criterion $\Tr[\rho\rho_{\rm prod}]^2/\Tr[\rho^2]\geqslant \Tr[\rho_{\rm prod}^2]-\beta^2$, where $\beta:=2^{1-m/2}\lambda_{\rm min}(\rho_{\rm prod})$, given in Ref.~\cite{23Wen_Ball}, which relied on the previous separable ball instead of the  larger ellipsoid given here. We note that Eq.~\eqref{eq:trace-condition} defines a convex space that strictly includes the ellipsoid centered at $\rho_{\rm prod}$, since it is equivalent to the set of $\rho$ that are made to satisfy the trace criterion around the identity via a SLOCC operation with filter $\mathcal F=(\Tr[\rho \rho_{\rm prod}]\rho_{\rm prod})^{-1/2}$. See Appendix~\ref{sec:convex-proof} for details.

\section{Generalized Criteria}
\subsection{Generalization to all separable states}
We now generalize the trace criterion \eqref{eq:trace-condition} to all separable states.

{\it Theorem 2 (Generalized Trace Criterion)}: Suppose a separable Hermitian matrix $K$ can be decomposed into the sum of positive product matrices, that is $K=\sum_{i=1}^u K_i$ where $K_i$ are positive (semi-)definite, product matrices such that $K_i=K_{i,1}\otimes...\otimes K_{i,m}$. If for a Hermitian matrix $\rho$, there exists an $i^*$ such that 3 conditions are met: i) $K_{i*}$ is full-rank; ii)
\begin{align}
\frac{\Tr[((\Delta+K_{i^*})K_{i^*}^{-1})^2]}{\Tr[(\Delta+K_{i^*})K_{i^*}^{-1}]^2} 
 \leqslant \frac{1}{D-c_m^2};
\label{eq:all-sep-condition}
\end{align}
where $\Delta \equiv \rho-K$ is such that iii) $\Delta+K_{i*}$ is positive definite, then $\rho$ is separable.

{\it Proof}: Suppose for some positive-definite product matrix $K_{i^*}$ where $1\leqslant i^*\leqslant u$, the inequality \eqref{eq:all-sep-condition} is reached with $\Delta K_{i^*}^{-1}+\mathbb{I}=(\Delta+K_{i^*})K_{i^*}^{-1}$. We have that $\Delta+K_{i^*}$ is separable according to Eq.~\eqref{eq:trace-condition}. Since
\begin{align}
\rho=K+\Delta=\left(\sum_{i\neq i^*} K_i\right)+(K_{i^*}+\Delta)\nonumber,
\end{align}
we have that $\rho$ is a mixture of separable matrices. \hfill\qedsymbol{}
 
The above theorem requires at least one of the components in the product-state decomposition of $K$ to be full-rank to calculate the inverse in Eq.~\eqref{eq:all-sep-condition}, which restricts $K$ to be full-rank. Using the generalized negative power defined for non-full-rank matrices, we can generalize \textit{Theorem 2} to all separable states (see Appendix~\ref{sec:non-full-rank} for details). 

We can use the above GTC to construct a hierarchy of separable states. Let us define $\mathcal S_u$ as the subset of separable states certified by the $u$-component GTC, i.e.,\/ the certifier $K$ is a mixture of at most $u$ product states (where $u$ is a positive integer). $\mathcal S_u$ possesses the following interesting properties. First, it is of full measure, since it contains the full-measure separable ellipsoid characterized by \textit{Theorem 1}. This is in contrast with the subset of separable states that can be expressed with a mixture of at most $u$ product states (including a full-rank one). In fact, $\mathcal S_u$ is a way to inflate this latter set into a full-measure one. Second, it is stable under SLOCC. Third, $\mathcal S_u$ is non-convex and strictly contains $\mathcal S_v$ for $v\leqslant u$. 
Fourth, it touches the boundary of separable states (SEP) at a number of points that increases with $u$ owing to the growing number of ways that the product states can become non-full rank. Finally, by increasing $u$ one can cover the full interior of SEP. The first sets in the hierarchy are schematically illustrated in (b) of Fig.~\ref{fig:ellipsoid}.

\subsection{Generalization to biseparability}
We next generalize \textit{Theorem 2} to the case of biseparable states, which have the form $  K_{\rm bisep} = \sum_{j=1}^u K_j $, where $K_j = K_{\mathcal{I}_j} \otimes K_{\overline{\mathcal{I}}_j}$ is a positive (semi-)definite product state for a bipartition $\mathcal{I}_j \cup \overline{\mathcal{I}}_j$ of the $m$ physical subsystems, and the sum runs over different bipartitions\footnote{The bipartitions $\mathcal{I}_j \cup \overline{\mathcal{I}}_j$ can be different for different indices $j$.} indexed by~$j$ \cite{01Acin,22Palazuelos}. We stress that biseparable states can be entangled, but do not possess any \textit{genuine} $m$-partite entanglement \cite{09Guhen_review}. Consider a state $\rho$ acting on the $m$ physical systems, and define $\Delta \equiv \rho-K_{\rm bisep}$. If for at least one $j^*$ the following 3 conditions are met: i) $K_{j*}$ is full-rank; ii)
\begin{equation}
     \frac{\Tr[((\Delta  + K_{{j^*}}) K_{{j^*}}^{-1})^2]}{\Tr[(\Delta  + K_{{j^*}}) K_{{j^*}}^{-1}]^2} \leqslant \frac{1}{D-c_2^2}\, ,
     \label{eq:bisep-criterion}
\end{equation} 
iii) $\Delta+K_{j*}$ is positive definite, then $\rho$ is biseparable.

The proof is direct. By applying Eq.~\eqref{eq:all-sep-condition} with $m=2$, we see that Eq.~\eqref{eq:bisep-criterion} implies that $\Delta+K_{j^*}$ is separable under the partition $\mathcal{I}_{j^*} \cup \overline{\mathcal{I}}_{j^*}$. We then observe that
\begin{equation}
    \rho = K_{\rm bisep}+\Delta =\left(\sum_{j\neq j^*} K_{j}\right) +(K_{j^*}+\Delta)\nonumber,
\end{equation}
namely $\rho$ has a biseparable form. 

\subsection{Separability detection}
To apply the criteria proposed in this work for showing the (bi)separability of an arbitrary state $\rho$, one needs to find a reference (bi)SEP state $K$ that can certify it. As a first step, one can apply Eq.~\eqref{eq:trace-condition} with the natural product state, the tensor product of the RDMs of each subsystem. 
However, many SEP states cannot be certified by such a criterion, for \emph{any} product state, so we turn to the GTC to obtain stronger results.

A direct procedure of applying the GTC would be to find among mixtures containing a fixed number of components $u$, $K=\sum_{i=1}^u K_i$, the one that is closest to $\rho$ by numerically minimizing the Hilbert-Schmidt distance $||\rho-K||_F$, and then using the GTC with this $K$. However, it turns out that states with the same minimal distance can lead to distinct minimal ratios in the GTC, making this procedure non-optimal.
We find that it is much more efficient to directly minimize the left-hand side of the GTC, Eq.~(\ref{eq:all-sep-condition}), over the set $\sum_{i=1}^u K_i$. Due to the widespread availability of optimization algorithms, this can be readily achieved, although finding the optimum becomes challenging when 1) the dimension increases, and 2) the state approaches the SEP boundary. 
In order to parametrize the $K_i$, we have found it convenient to employ a Cholesky decomposition $K_i=LL^\dag$, where $L$ is lower triangular. Furthermore, in using the GTC, we imposed $\Tr[K]=1$, which we found led to valid certification of separability, i.e.,\/ a $K$ satisfying $\Delta+K_{i*}>0$.

\section{Applications}
As a warmup, we first  apply the ellipsoid and GTC criteria on 3-qubit X-states in a dephasing environment \cite{weinstein2010entanglement}. A generic 3-qubit X state $\rho_X$ depends on three sequences of four parameters $\boldsymbol{a},\boldsymbol{b},\boldsymbol{c}$ where $\boldsymbol{a} = \{a_1,a_2,a_3,a_4\}$, and has an X-shape in the computational basis (see Appendix~\ref{sec:three-qubit-X}). In the presence of an independent qubit dephasing environment with parameter $0\leqslant p \leqslant 1$, the antidiagonal $c_j$ coefficients are  multiplied by a factor $(1-p)^{3/2}$~\cite{weinstein2010entanglement}. The dephased state is $\rho(p) \equiv\rho_X(\boldsymbol{a},\boldsymbol{b},(1-p)^{3/2}\boldsymbol{c})$ leading to a separable state at $p\!=\!1$. 
Let us first test Eq.~\eqref{eq:trace-condition} with the natural product state $\rho_{\rm prod} = \rho_1 \otimes \rho_2 \otimes \rho_3$, where $\rho_1 = \Tr_{2,3}[\rho_X]$ is the RDM for the first qubit, and similarly for $\rho_2,\rho_3$. As an example, we choose $\boldsymbol{a} = \{\frac{1}{8}, \frac 18, \frac{1}{32},\frac{1}{64}\}$, $\boldsymbol{b} = \{\frac{1}{8},\frac{1}{8},\frac{7}{32},\frac{15}{64}\}$ and $\boldsymbol{c} = \{\frac{1}{12}, \frac{1}{24}, \frac{1}{24},\frac{1}{36}\}$. With these parameters, the ratio of volumes between the ellipsoid and the ball centered on $\rho_{\rm prod}$ reads $\mathcal{R} \approx 10^{24}$, using Eq.~\eqref{eq:R}. With the ellipsoid criterion of Eq.~\eqref{eq:ellipsoid-condition} centered on $\rho_{\rm prod}$, we find that the dephased state $\rho(p)$ is separable for $p\geqslant 0.5$, whereas the ball criteria of Ref.~\cite{23Wen_Ball} never detects separability, even for $p=1$. For $p \geqslant 0.5$, the dephased state lies within the ellipsoid centered on $\rho_{\rm prod}$, but is never included in the separable ball. This example illustrates that our ellipsoid criterion detects more separable states compared to the ball criterion. Moreover, with the trace criterion of Eq.~\eqref{eq:trace-condition}, we find that the dephased state is separable for $p\geqslant 0.47$, verifying that the simple trace criterion is indeed stronger. We then employ the full force of the GTC: using separable states constructed as a sum of 12 arbitrary product states, we readily show that $\rho(p)$ becomes separable at $p=0.1937$, which is extremely close to the PPT threshold, $p_{\rm PPT}=0.193$. We provide strong evidence for the PPT condition being necessary and sufficient for the full separability of the X-state under consideration.
The same methods can be applied to three-qubit X states in a depolarizing environment~\cite{weinstein2010entanglement}, and we again find examples of states which lie in the ellipsoid but not in the separable ball centered on the natural product state. 

\subsection{Benchmarks} 
First, we benchmark the GTC with a robustness test on two pure 4-qubit states: $|W_4\rangle$ \cite{02Verstraete} and the Higuchi-Sudbery (HS) state \cite{Higuchi2000}. We determine the value of white noise necessary to make $(1-p)|\psi\rangle\langle\psi| +p \mathbb{I}/D$ separable.  In order to improve the performance of the algorithm, we iterate as follows: optimizing the GTC over $S_{u_1}$, get a valid output state $K$. If it does not certify separability, use it to construct a new state $K^{-1/2}_{i*}(\rho-K+K_{i*})K^{-1/2}_{i*}$, normalize it to get $\rho'$, and feed it into the GTC optimized over $\mathcal S_{u_2}$. This leads to a $K'$ with a lower ratio of the LHS/RHS in the GTC. The iteration is to be continued until certification is achieved.

For $W_4$, an exact separability threshold is known, which the GTC nearly matches. As a comparison, it outperforms the iterative algorithm of Kampermann \emph{et al}~\cite{Kampermann2012} (last column in Table~\ref{tab:benchmarks}) and an approach based on neural networks~\cite{neural2022}.
For the HS state~\cite{Higuchi2000}, the exact robustness is not known, but the GTC approaches the PPT threshold. 
Interestingly, the HS state is less robust than $W_4$, and our result obtained with the iterative algorithm~\cite{Kampermann2012} shows that PPT essentially determines the separability threshold.
The above benchmarks show that the GTC can produce strong separability results.

\subsection{Quantum Ising} 
As an important physical application, we now consider the 1d transverse field Ising model at the critical coupling $h=1$ and  finite temperature described by the Gibbs state $\exp(-\beta H)/\mathcal Z$, where $\beta=1/T$ is the inverse temperature and $H$ is given in Eq.~\eqref{eq:1D-Ising}. We study the RDM $\rho(T)$ of three and four adjacent sites in an infinite chain, which can be obtained exactly (see Ref.~\cite{fagotti2013reduced} and references therein). The results are shown in Table~\ref{tab:benchmarks}. For 3 spins the GTC gets extremely close to the PPT threshold for both full-separability and separability with respect to the bipartition $1|23$, leaving little room for further improvement. 
Interestingly, it follows that the density matrix has no bipartite bound entanglement with respect to the partition $1|23$, or tripartite bound entanglement. In the latter case, it means that there is no temperature regime where the state is PPT with respect to every bipartition but not fully separable. For 4 spins, we get close to the PPT threshold. 
Finally, we apply the biseparable version of the GTC on 3 spins and find that the state becomes biseparable for $T\geqslant 0.87$, a value comparable to the one obtained by the iterative algorithm.

\begin{table}
    \centering
    \begin{tabular}{l|c|c|cc}
    \hline
    \multirow{2}{*}{State} & \multirow{2}{*}{Parameter} & \multirow{2}{*}{Lower Bound} & \multicolumn{2}{c}{Upper Bounds} \\
    \cline{4-5}
          &           &              & GTC & K \\
    \hline
    \multirow{2}{*}{$\rho_{3,\text{Ising}}$} & $T_{\text{s},1\vert23}$ & $1.1634^{\sharp}$  & 1.1636  & 1.166  \\ 
    & $T_{\text{s}}$ & $1.3637^{\sharp}$  & 1.3639  & 1.370  \\
     & $T_{\text{bs}}$ & $0.74^{\star}$  & 0.87  & 0.84  \\ 
     \hline
    $\rho_{4,\text{Ising}}$ & $T_{\text{s}}$ & $1.497^{\sharp}$   & 1.54    & 1.51  \\
    \hline
    $W_4$                  & $p_{\text{s}}$ & $0.9074^{*}$\cite{Zhou2022}  & 0.9095    & 0.91\cite{Kampermann2012}  \\
    \hline
    HS & $p_{\text{s}}$ & $8/9\approx 0.\bar{8}^\sharp$  & 0.909 & 0.890  \\
    \hline
    \end{tabular}
    \caption{Bounds for (bi)separability parameters. For the 1d Ising RDM at its critical field \(h=1\), the fully separable temperature \(T_s\) is reported for three- and four-site adjacent clusters (\(\rho_{3,\text{Ising}}\) and \(\rho_{4,\text{Ising}}\)), while the separable temperature for partition 1$\big|$23 and the biseparable temperature \(T_{bs}\) is provided for the three-site state. For the four-qubit \(W_4\) state and the Higuchi–Sudbery (HS) state~\cite{Higuchi2000}, the separable probability threshold \(p_s\) is given. Lower bounds are obtained via the PPT criterion (denoted \(^{\sharp}\)), genuine multipartite negativity (\(^\star\))~\cite{Jungnitsch2011} or from literature (\(^{*}\)), while upper bounds are derived using the Generalized Trace Criterion (GTC, Eq.~\eqref{eq:all-sep-condition}) and the iterative algorithm of Kampermann \emph{et al} (K)~\cite{Kampermann2012}.}
    \label{tab:benchmarks}
\end{table}

\section{Conclusion} 
In this work, we have generalized the previous separability conditions by identifying an ellipsoid of separable states centered around any finite-dimensional, $m$-partite product state. Such ellipsoid contains and significantly improves the separable ball around any product state~\cite{20Hanson_Markovian,23Wen_Ball}: the volume of the ellipsoid is much larger than that of the previous ball. We then further improved the ellipsoid around the product state by applying a scaling coefficient to find a trace condition Eq.~\eqref{eq:trace-condition}, leading to our main result, the GTC in Eq.~\eqref{eq:all-sep-condition} and its biseparable version Eq.~\eqref{eq:bisep-criterion}, which characterize a (bi)separable region around any 
 (bi)separable state. The GTC gives rise to a hierarchy of non-convex subsets that cover the entirety of the separable set. We then showed that the GTC can be numerically used to obtain strong separability and biseparability results for 3- and 4-qubit states, including in the finite-temperature 1d quantum Ising model. We expect our criteria will have wide applicability in detecting separability.

\section*{Acknowledgements} 
A.K. acknowledges support through a Discovery Grant of the National Science and Engineering Council of Canada (NSERC), an Applied Quantum Computing Challenge Grant of the National Research Council (NRC) of Canada, and a Discovery Project grant of the Australian Research Council (ARC). GP held FRQNT and CRM-ISM postdoctoral fellowships, and received support from the Mathematical Physics Laboratory of the CRM while this work was carried out. W.W.-K.\ and L.L.\/ are supported by a grant from the Fondation Courtois, a Chair of the Institut Courtois, a Discovery Grant from NSERC, and a Canada Research
Chair. RYW acknowledges support through the Canada Graduate Research Scholarship – Doctoral program (CGRS D) from NSERC.

\appendix
\section{Criterion for non-full-rank states} \label{sec:non-full-rank}

Since we have the ellipsoid characterized by Eq.~\eqref{eq:ellipsoid-condition} in the main text to capture the directional dependence of the separable states around the product state, we can now take into account the non-full-rank product state by restricting the transformation where the eigenvalues do not vanish. Consider a Hermitian matrix (not necessarily full-rank) $A=\sum_{i}^{D_{\rm f}}a_i\ket{a_i}\bra{a_i}$ where $a_i\neq 0$ and $D_{\rm f}={\rm rank}(A)\leqslant{\rm dim}(A)=D$. Define $P_{\rm f}$ as the projector to the full-rank subspace of $A$ and $P_{\rm n}:=\mathbb{I}-P_{\rm f}$ as the complement to $P_{\rm f}$. We denote $X_{{\rm f}}:=P_{\rm f}XP_{\rm f}$ to be the projection of any matrix $X$ into the full-rank subspace of $A$. We then define the generalized negative power of $A$ to $-p$ (with $p$ positive) as: $A^{(-p)}:=\sum_{i}^{D_{\rm f}}1/a_i^p\ket{a_i}\bra{a_i}$. With these definitions, we can easily generalize {\it Theorem 1} to the case of a non-full-rank $\rho_{\rm prod}$:

{\it Corollary 1}: Consider a positive (semi-)definite Hermitian operator $\rho_{\rm prod}:=\rho_1\ot...\ot \rho_m$. If a Hermitian operator $\rho$ satisfies $\lVert \rho_{\rm prod}^{(-1/2)}\rho\rho_{\rm prod}^{(-1/2)}-\mathbb{I}_{\rm f}\lVert_{\rm F}\leqslant c_m$ and $\rho=\rho_{\rm f}$, then $\rho$ is separable. 

We can prove the above corollary similarly to {\it Theorem 1}. We essentially require $\rho$ to only reside and to satisfy the separability criterion of {\it Theorem 1} in the full-rank subspace of $\rho_{\rm prod}$. Therefore, the separable region around the non-full-rank product state is effectively a lower dimensional ellipsoid. 
The direction of the lower dimensional ellipsoid is consistent with what we know for non-full-rank product states: arbitrarily weak perturbations in the vanishing subspace of the non-full-rank $\rho_{\rm prod}$ can make the state entangled \cite{16LamiEntanglementSaving,22Wen}. 

Similar to Eq.~\eqref{eq:trace-condition} for the full-rank product states, we can improve {\it Corollary 1} to a trace condition that applies to any product states $\rho_{\rm prod}$ with $D_{\rm f}={\rm rank}(\rho_{\rm prod})$:

{\it Corollary 2}: For any Hermitian $\rho$ satisfying $\rho=\rho_{{\rm f}}$ and 
\begin{equation}
\frac{\Tr[(\rho\rho_{\rm prod}^{(-1)})^2]}{\Tr[\rho\rho_{\rm prod}^{(-1)}]^2} \leqslant \frac{1}{D_{\rm f}-c_m^2},
\label{eq:trace-condition-2} 
\end{equation}
we have that $\rho$ is separable.

Using the above {\it Corollary 2}, we can generalize {\it Theorem 2} in the main text to all separable states, without the previous restriction of $K$ having at least one full-rank component in its decomposition:

{\it Corollary 3}:  Consider any separable Hermitian matrix $K$ which can be decomposed into the sum of product states, that is $K=\sum_{i=1}^u K_i$ where $K_i$ are positive (semi-)definite, product matrices. Let $X_{{\rm f}_i}$ be the projection of any matrix $X$ in the full-rank subspace of $K_i$ with dimension $D_i$. If for a Hermitian matrix $\rho$ there exists some $i=i^*$ such that i) $\Delta_{{\rm f}_{i^*}}=\Delta$ where $\Delta \equiv \rho-K$; ii)
\begin{align}
\frac{\Tr[(\Delta K_{i^*}^{(-1)}+\mathbb{I}_{{\rm f}_{i^*}})^2]}{\Tr[\Delta K_{i^*}^{(-1)}+\mathbb{I}_{{\rm f}_{i^*}}]^2}\leqslant \frac{1}{D_{i^*}-c_m^2}\,;
\label{eq:all-sep-condition-any}
\end{align}
and iii) $\Delta+K_{i*}$ is positive (semi-)definite, then $\rho$ is separable.

The proof for the above criterion is the same as {\it Theorem 2} except using Eq.~\eqref{eq:trace-condition-2} instead of Eq.~\eqref{eq:trace-condition}, which allows $K_{i^*}$ to be non-full-rank. To use Eq.~\eqref{eq:all-sep-condition-any}, one has to diagonalize each $K_{i}$ of the product-state decomposition of $K$, use its eigenvectors to calculate its generalized inverse $K_{i}^{(-1)}$, and then check whether the inequality in Eq.~\eqref{eq:trace-condition-2} is satisfied. Eq.~\eqref{eq:all-sep-condition-any} generalizes {\it Theorem 2} to all separable states and forms a necessary and sufficient condition for separability, but it is much harder to use in practice. Numerically, it is difficult to find a decomposition of $K$ such that one of its component satisfies $\Delta_{{\rm f}_{i^*}}=\Delta$ as required in {\it Corollary 3}. We can also use Eq.~\eqref{eq:all-sep-condition-any} to obtain a similar criterion for bisparable states, which will generalize Eq.~\eqref{eq:bisep-criterion} in the main text.

\section{Three-qubit X state}\label{sec:three-qubit-X}
The three-qubit X state is
\begin{equation}
    \rho_X(\boldsymbol{a},\boldsymbol{b},\boldsymbol{c}) = \begin{pmatrix}
        a_1 & 0 & 0& 0& 0& 0& 0& c_1 \\
        0&a_2&0&0&0&0&c_2&0\\
        0&0&a_3&0&0&c_3&0&0\\
         0&0&0&a_4&c_4&0&0&0\\
         
         0&0&0&c_4^*&b_4&0&0&0\\
         0&0&c_3^*&0&0&b_3&0&0\\
         0&c_2^*&0&0&0&0&b_2&0\\
          c_1^* & 0 & 0& 0& 0& 0& 0& b_1 
    \end{pmatrix},
    \label{eq:three-qubit-X}
\end{equation}
and the one-qubit reduced density matrices read 
\begin{equation}
\begin{split}
    \rho_1 &= \begin{pmatrix}
        a_1+a_2+a_3+a_4 & 0 \\
        0 & b_1+b_2+b_3+b_4
    \end{pmatrix},\\
     \rho_2 &= \begin{pmatrix}
        a_1+a_2+b_3+b_4 & 0 \\
        0 & b_1+b_2+a_3+a_4
    \end{pmatrix},\\
     \rho_3 &= \begin{pmatrix}
        a_1+b_2+a_3+b_4 & 0 \\
        0 & b_1+a_2+b_3+a_4
    \end{pmatrix}.
    \end{split}
     \label{eq:three-qubit-X-reduced}
\end{equation}
In particular, these reduced density matrices do not depend on the off-diagonal $\boldsymbol{c}$ elements. 

\section{Convexity of the Trace criterion}\label{sec:convex-proof}

We wish to show that the set of density matrices satisfying 
\[
\frac{\operatorname{Tr}\left[(\rho\, \rho_{\mathrm{prod}}^{-1})^2\right]}{\operatorname{Tr}\left[\rho\, \rho_{\mathrm{prod}}^{-1}\right]^2} \leqslant \frac{1}{D-c_m^2}
\]
is convex. To this end, we reformulate the criterion using a filtering operation. Given a full-rank product state \(\rho_{\mathrm{prod}}\), define the unnormalized filtering operation as \(\mathcal{F}[\rho] = \rho_{\mathrm{prod}}^{-1/2}\,\rho\,\rho_{\mathrm{prod}}^{-1/2}\) and let \(N = \operatorname{Tr}(\mathcal{F}[\rho])\), so that the normalized filtered state is \(\tilde{\rho} = \mathcal{F}[\rho]/N\). The trace criterion is equivalent to requiring that \(\tilde{\rho}\) lies within the separable ball defined by \(\operatorname{Tr}(\tilde{\rho}^2) \leqslant \frac{1}{D-c_m^2}\).

Assume that two density matrices \(\rho_1\) and \(\rho_2\) satisfy the trace criterion, and let their normalized filtered states be \(\tilde{\rho}_1 = \mathcal{F}[\rho_1]/N_1\) and \(\tilde{\rho}_2 = \mathcal{F}[\rho_2]/N_2\) with \(N_i = \operatorname{Tr}(\mathcal{F}[\rho_i])\) for \(i=1,2\). Consider a convex combination \(\rho = p\,\rho_1 + (1-p)\,\rho_2\) with \(0 \leqslant p \leqslant 1\). By linearity, \(\mathcal{F}[\rho] = p\,\mathcal{F}[\rho_1] + (1-p)\,\mathcal{F}[\rho_2]\) and \(N = p\,N_1 + (1-p)\,N_2\), so that
\[
\tilde{\rho} = \frac{\mathcal{F}[\rho]}{N} = \frac{p\,N_1}{N}\,\tilde{\rho}_1 + \frac{(1-p)\,N_2}{N}\,\tilde{\rho}_2.
\]
Since the weights \(\frac{p\,N_1}{N}\) and \(\frac{(1-p)\,N_2}{N}\) are nonnegative and sum to one, \(\tilde{\rho}\) is a convex combination of \(\tilde{\rho}_1\) and \(\tilde{\rho}_2\). Given that the separable ball is convex, it follows that \(\tilde{\rho}\) also lies in the separable ball, and hence the original state \(\rho\) satisfies the trace criterion. This completes the proof that the set defined by the trace criterion is convex.

\bibliography{refs2}

\end{document}